\documentclass[manuscript]{aastex}
\shorttitle{Diffusion in sectored field}
\shortauthors{Florinski et al.}

\begin{document}

\title{Cosmic-ray diffusion in a sectored magnetic field in the distant
heliosheath}

\author{V. Florinski\altaffilmark{1,2}, F. Alouani-Bibi\altaffilmark{2},
J. Kota\altaffilmark{3}, and X. Guo\altaffilmark{2}}

\altaffiltext{1}
{Department of Physics, University of Alabama, Huntsville, AL 35899}

\altaffiltext{2}{Center for Space Plasma and Aeronomic Research, University of
Alabama, Huntsville, AL 35899}

\altaffiltext{3}{Lunar and Planetary Laboratory, University of Arizona, Tucson,
AZ 85721}

\begin{abstract}
Very high intensities of galactic cosmic rays measured by Voyager 1 in the
heliosheath appear to be incompatible with the presence of a modulation ``wall''
near the heliopause produced by a pile up of the heliospheric magnetic field.
We propose that the modulation wall is a structure permeable to cosmic rays as
a result of a sectored magnetic field topology compressed by plasma slowdown on
approach to the heliopause and stretched to high latitudes by latitudinal flows
in the heliosheath.
The tightly folded warped current sheet permits efficient cosmic-ray transport
in the radial direction via a drift-like mechanism.
We show that when stochastic variations in the sector widths are taken into
account, particle transport becomes predominantly diffusive both along and
across the magnetic sectors.
Using a test-particle model for cosmic rays in the heliosheath we investigate
the dependence of the diffusion coefficients on the properties of the sector
structure and on particle energy.
\end{abstract}


\section{Introduction}

The twin Voyager space probes are currently exploring the distant heliosheath
region and are expected to reach the heliospheric boundary, known as the
heliopause, and pass into the surrounding interstellar space before the end of
the decade.
Before crossing the heliopause, the Voyagers are expected to enter a region of
rapidly changing magnetic field which is a remnant of magnetic sectors separated
by an oscillating heliospheric current sheet (HCS).
From global heliospheric simulations reported by \citet{czechowski10},
\citet{borovikov11}, and \citet{florinski11}, it emerges that sectors become
narrower as the solar wind plasma slows down as it nears the heliopause.
It was suggested by \citet{florinski11} that highly energetic charged particles,
such as galactic cosmic ray ions, will be profoundly affected by the compressed
sector structure in the heliosheath.
Charged particles find it much easier to travel across the sectors, and hence
across the magnetic field, when the width of a sector is comparable to or
smaller than their Larmor radius (rather than being much larger than the latter,
as is typically the case in the solar wind).
The enhanced cross-field transport process is a consequence of fast drift-like
motion across the stack of magnetic sectors.

Voyager observations \citep{burlaga03} show no regular magnetic sector pattern
in the distant solar wind, a trend that appears to persist into the heliosheath
\citep{burlaga10}.
Measured sector durations in the solar wind were, on average, shorter than 13
days expected from a simple two-sector pattern implying a significant distortion
of the shape of the HCS.
The widths of magnetic sectors reported by \citet{burlaga03} varied by almost
a factor of 10 and had a sufficiently broad distribution.

\citet{burlaga03} only reported sectors observed during three years near the
maximum of solar activity.
Close to solar minima magnetic sectors occupy only a narrow range of
heliolatitudes.
They were not observed by the Voyagers during the pre-termination shock part of
the mission when the spacecraft was above (V1) or below (V2) the current sheet.
The situation changes dramatically in the heliosheath.
Here the flow is deflected away from the radial direction \citep{richardson09}
stretching the HCS in latitude and longitude \citep{czechowski10, borovikov11}.
In the hemisphere facing the interstellar flow the sector structure is expected
to exist at all latitudes deep into the heliosheath.
It is then of great importance to understand the effect of the tightly folded
current sheet on energetic particles, such as galactic cosmic rays.

It was proposed in \citet{florinski11} that variability in sector width results
in an enhanced cross-field diffusion of galactic cosmic rays with a Larmor
radius greater than about 0.1 AU.
The process could well be responsible for very high intensities of galactic
helium ions measured by Voyager 1, intensities that are presently (2011--2012)
near or in excess of the predicted interstellar values \citep{webber09}.
\citet{florinski11} used a simple, one-dimensional sector model that did not
take into account variations in the shape of current sheets.
The present paper reports a much more complete, two-dimensional model of
particle transport near a rapidly oscillating HCS.
Here we assume that the sector width in the distant heliosheath varies
about a well defined mean, and that the variation occurs as a result of
differential (variable in latitude) current sheet displacement in the radial
direction.
This work therefore builds upon the ideas proposed in \citet{florinski11} to
develop a \textit{quantitative} model of cosmic-ray transport in a sectored
magnetic field of the distant heliosheath.

It is noteworthy that a partially related concept was proposed much earlier as a
mechanism of interstellar dust grain transport in the solar wind
\citep{morfill79, czechowski03}.
High rigidity and small velocity of dust grains makes their transport
qualitatively different from that of cosmic rays.
Dust particles larger than about 0.01 $\mu$m have gyroradii comparable to the
size of the heliosphere and their trajectories are only mildly perturbed by
the sector structure in the solar wind.
These slowly moving particles acquire a velocity component perpendicular to the
interplanetary magnetic field and parallel to the sectors as a result of the
motional electric field in the frame of the particle.
Random variations in sector duration then lead to a diffusion in pitch angle.
For galactic cosmic rays with their high velocities this effect is negligible,
and the principal mode of transport is drift associated with the current sheet
crossings.

The paper is structured as follows.
In Section 2 we discuss the fundamental physics of particle drift in a sectored
magnetic field.
In section 3 we investigate cosmic ray transport through a stack of magnetic
sectors with a test-particle orbit integration method commonly used to
simulate scattering and cross-field transport in turbulent magnetic fields
\citep[e.g.,][]{giacalone99, mace00, qin02}.
We demonstrate that particle transport in a magnetic field composed of
variable-width sectors of alternating polarity is diffusive both parallel to
and across the current sheets.
The results from this study are reported in Section 4.
These results reveal the dependence of the diffusion coefficients on the
distribution of sector widths and on the cosmic-ray energy.
Finally, Section 5 explores some consequences of the new diffusion mechanism for
cosmic-ray propagation through the heliosheath.

\section{Neutral sheet diffusion as a drift process}
Consider an energetic charged particle, such as a galactic cosmic ray ion placed
in a magnetic field that has the same magnitude everywhere, but can change sign.
A change of sign occurs across a neutral sheet with negligible thickness (i.e.,
thickness much smaller than the Larmor radius of the energetic particle).
It is well known that in the absence of a large-scale plasma flow, the particle
can either gyrate in place (ignoring the motion along the magnetic field, which
is of no interest here), or undergo a meandering motion (drift) along a single
current sheet.
When the sectors become squeezed together, as one expects will happen on
approach to the heliopause, the distance between adjacent neutral sheet folds
would become comparable or smaller than the cosmic ray cyclotron radius $r_g$,
which is typically between 0.05 and 0.3 AU in the heliosheath.

\begin{figure}
\plotone{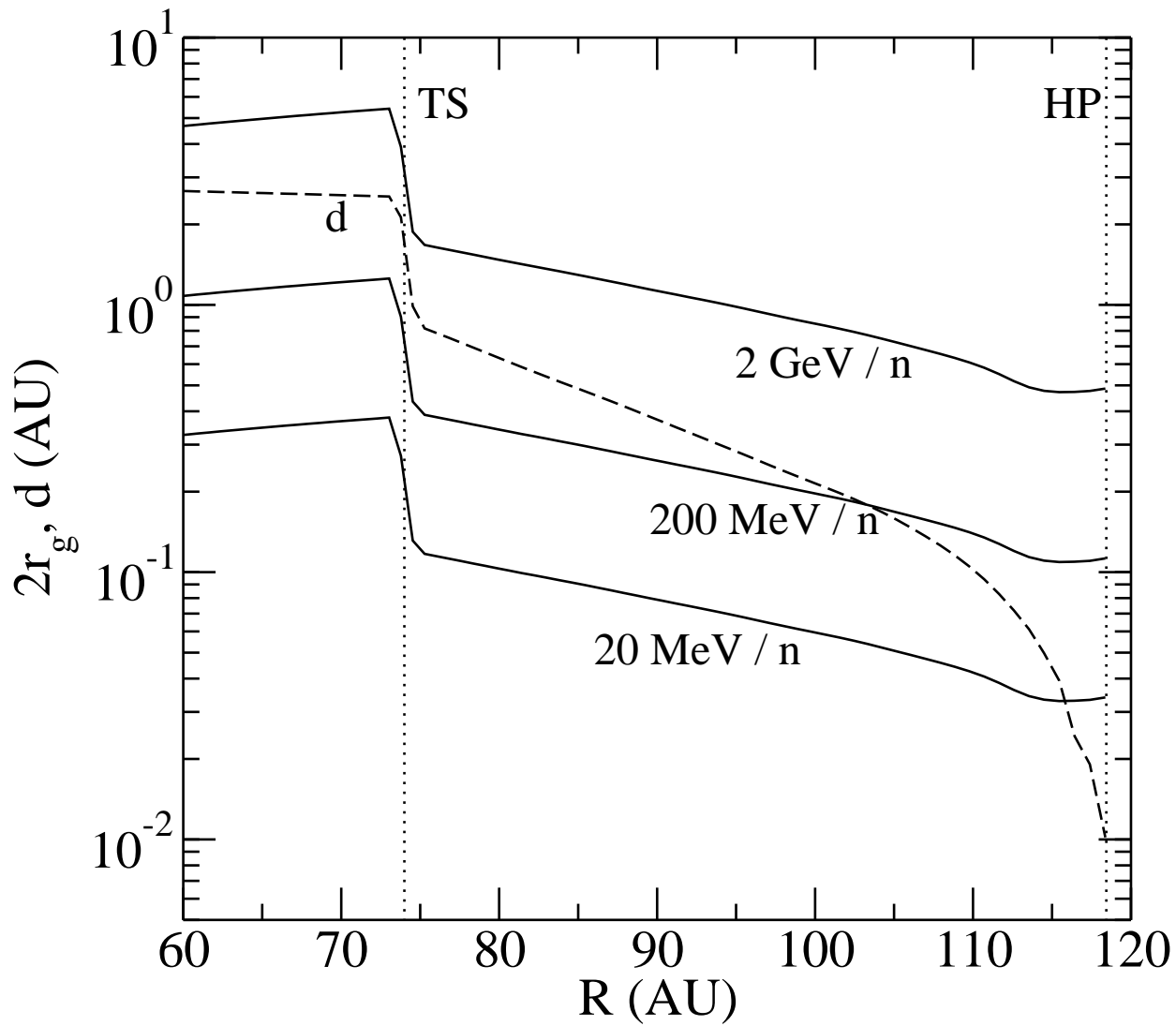}
\caption{Larmor radius of galactic helium at three different energies per
nucleon (solid lines) and the mean thickness of a magnetic sector in the
heliosheath in the upwind (nose) direction.
Current sheet diffusion is expected to be important between the intersection
point of the $2r_g$ and $d$ curves and the heliopause.
The positions of the termination shock and the heliopause are marked with dotted
lines.
\label{fig_radial}}
\end{figure}

The slowly moving heliosheath plasma is thought to be nearly incompressible,
which appears to be consistent with Voyager 2 results \citep{richardson11}.
The plasma is therefore not compressed, but follows different paths toward the
tail (spreads out in latitude and longitude).
Evidently, for primarily north-south flows the magnitude of the mean field need
not increase because the decrease in the radial distance between field lines
(as the sectors become thinner) is balanced by spreading of the field lines in
latitude.
For azimuthal flows, the magnitude of $\mathbf{B}$ should increase because in
that case plasma follows magnetic field lines.
Altogether, we expect the magnetic field to increase by a modest amount, so that
the ratio of $r_g$ to the mean distance between sectors $d$ increases with
heliocentric distance.
At some distance, therefore, the critical condition $2r_g=d$ will be satisfied.

To support our estimate we calculated the magnetic field and the mean sector
width in the heliosheath as a function of radial distance from the Sun using a
three-dimensional model of the heliosphere that includes neutral hydrogen atoms
\citep{florinski11}.
From this data we calculated the Larmor radius $r_g$ for 20, 200, and 2000 MeV/n
He$^{2+}$ ions as a function of heliocentric distance.
Figure \ref{fig_radial} shows the radial dependence of $d$ and $2r_g$ in the
upwind (nose) direction.
Because of its blunt shape, the termination shock stand-off distance, marked by
the dotted line, is smaller ($\sim 74$ AU) than in the directions of Voyager 1
and 2.
The model is steady-state and does not include dynamic solar wind structures,
but is sufficient to illustrate our point.
Readers interested in the model boundary conditions for the solar wind and the
local interstellar medium are referred to \citet{florinski11}.

From the figure, the gyro-radius does decrease with distance, but not as
rapidly as the sector width.
In the model the magnetic field magnitude increases from 0.15 nT to 0.5 nT
between the termination shock and the heliopause, whereas the sector width
decreases by nearly two orders of magnitude.
The point of intersection between the solid and the dashed curves in Figure
\ref{fig_radial} ($2r_g=d$) gives the approximate distance beyond which the
effects of neutral shift diffusion become noticeable.
For 200 MeV/n ions this occurs about two-thirds of the way through, but only
close to the heliopause for 20 MeV/n and lower energy particles.

\begin{figure}
\plotone{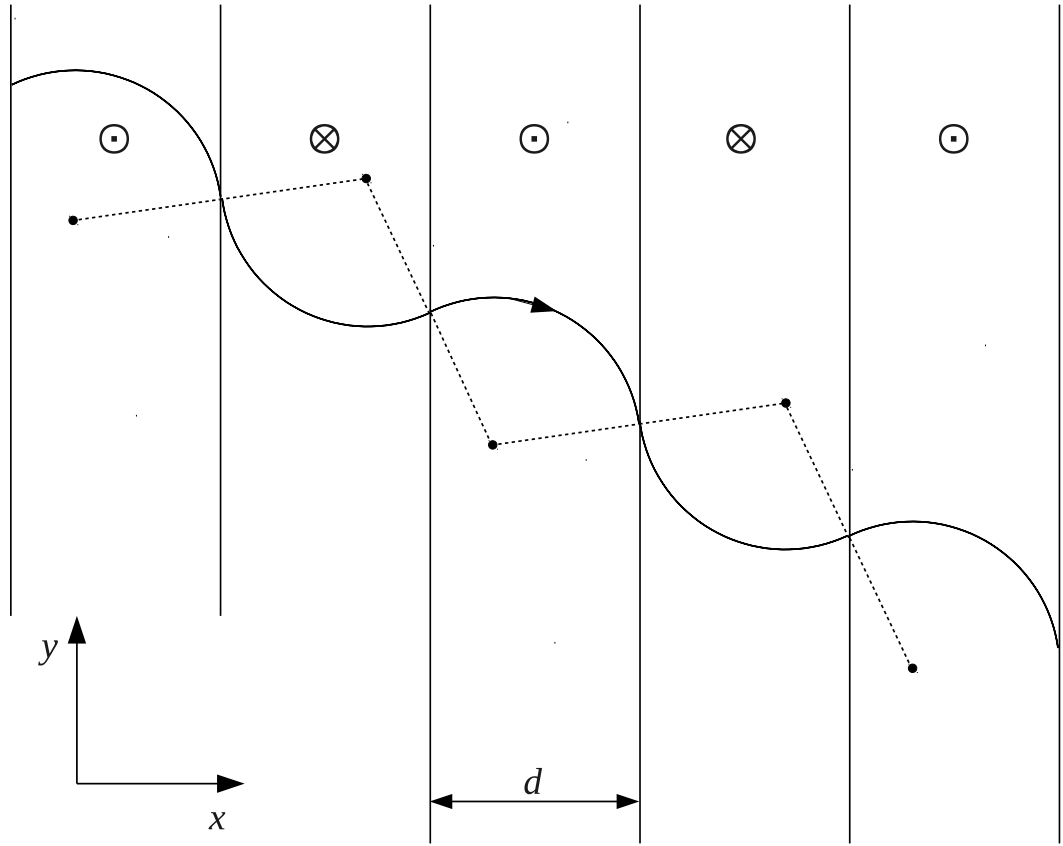}
\caption{Drift motion of a positively charged ion through a stack of magnetic
sectors of alternating polarity.
The solid line is the particle's trajectory and the dotted line shows the
``trajectory'' of its guiding center.
The magnetic field is in and out of the plane of the figure.
\label{fig_drift}}
\end{figure}

As discussed in \citet{florinski11}, for $r_g\leq d/2$ there exist gyrophase
angles for which meandering or gyrating particles are unable to complete their
cycle before encountering a neighboring current sheet.
At that point the particle's sense of rotation will change as its guiding center
``jumps'' across the sheet and into an adjacent sector.
Figure \ref{fig_drift} shows a typical trajectory of an energetic particle in an
ensemble of sectors of uniform width.
The particle drifts in both $y$ and $x$ directions.
However, unlike the former drift, it is easy to see by drawing a similar diagram 
for a different starting gyrophase, that drift across the magnetic boundaries
can proceed in either direction (to the left or to the right on the figure,
depending on the gyrophase).
It follows that the net drift velocity is zero for a gyrotropic ensemble of
particles.
However, the process itself may lead to a separation of an initially symmetric
distribution into left- and right-going components.

Despite the limitation, the drift process discussed above could enhance the
\textit{diffusive} transport of energetic particles if the sector structure is
not uniform.
Observations clearly show that the sector widths (or duration, in terms of
spacecraft measurements) vary by a factor of $\sim 10$ in the outer
heliosphere \citep{burlaga03}.
This variability is presumably due to both an irregular magnetic field structure
on the Sun and the dynamic interaction between structures in the solar wind
further distorting the shape of the HCS \citep[e.g.][]{riley02, intriligator05}.
One expects the sector thickness to vary with both latitude and longitude; we
include only latitudinal variations in our two-dimensional model (see below).
The advantage of the present model is simplicity stemming from the assumption
that the magnetic field varies in direction only, but not in magnitude (this
condition cannot be maintained in the fully three-dimensional case).
This allows us to focus on the fundamental physics of the diffusion process due
to current sheet fluctuations, a feature that would otherwise be hidden
underneath the complexity of a fully 3D simulation.

In the last part of this section we present a simple two-dimensional model,
based on the Boltzmann equation, that illustrates the qualitative features
expected from closely packed alternating sectors.
In this model, the sector width is fixed and the randomness is provided by
scattering centers embedded in the plasma.
For the sake of brevity, we consider only the motion in the $xy$ plane (with the
background magnetic field directed along $\hat\mathbf{z}$), and disregard
scattering in pitch angle, which would lead to quantitative modifications but
would not change the robust qualitative features.
We assume the pitch angle to be $\pi/2$ and consider scattering in gyrophase
angle $\varphi$ only.

As in Figure \ref{fig_drift} we take the distribution function of cosmic rays to
be independent of $y$.
Assuming that diffusion is a result of independent small-angle scattering in
$\varphi$, the Boltzmann equation for the phase space density $f(x,\varphi ,t)$
can be written as
\begin{equation}
\frac{\partial f}{\partial t}+v\cos\varphi\frac{\partial f}{\partial x}
-\Omega\frac{\partial f}{\partial\varphi}
=\frac{\partial}{\partial \varphi}
\left(\frac{1}{\tau}\frac{\partial f}{\partial\varphi}\right),
\end{equation}
where $\Omega = v/r_g$ is the gyro-frequency and $\tau$ is the mean scattering
time.
The velocity $v$ is the same for all particles.
Physically, $\tau^{-1}$ is related to the quasi-linear Fokker-Planck coefficient
$D_{\varphi\varphi}$ that describes diffusion in gyrophase through resonant
interactions with turbulent magnetic fields \citep{achatz91}.
The latter is a function of pitch-angle cosine $\mu$ and velocity $v$; since
both are constant in our model, we accordingly take $\tau=\mathrm{const}$.

For a unipolar field pointing in the $\hat\mathbf{z}$ direction (i.e. constant
$\Omega$), Eq. (1) entertains the following steady-state solution 
\begin{equation}
f(x,\varphi)=F_0+G_x x  
+\frac{2 S^0_x}{v}\cos\varphi+\frac{2 S^0_y}{v}\sin\varphi,
\end{equation}
implying that a steady gradient ($G_x$) produces streaming
$S^0_{x,y}=(2\pi)^{-1}\int f v_{x,y}d\varphi=-\kappa_{(x,y)x}G_x$ in the
$\hat\mathbf{x}$ and $\hat\mathbf{y}$ directions, respectively, consistently
with the nominal diffusion coefficients of
$\kappa_{xx}=v^2\tau/[2(1+\Omega^2\tau^2)]$ and
$\kappa_{yx}=-\Omega\tau\kappa_{xx}$.

Next, we place alternating sectors of widths $d_+$ and $d_-$ periodically 
in the $yz$ plane. 
Though the solution shown in Eq. (2) remains formally a solution, one should 
bear in mind that $S_y$ now becomes discontinuous at the current sheets, 
while the physical solution must be continuous.
We note that an approach based on Parker's equation \citep{parker65} cannot be
directly applied near current sheets because Parker's equation does not include
shear.
Its extension to include second-order terms shows that shear produces
significant higher harmonics \citep{kota75, earl88}, which cannot be neglected
in our case.

\begin{figure}
\plotone{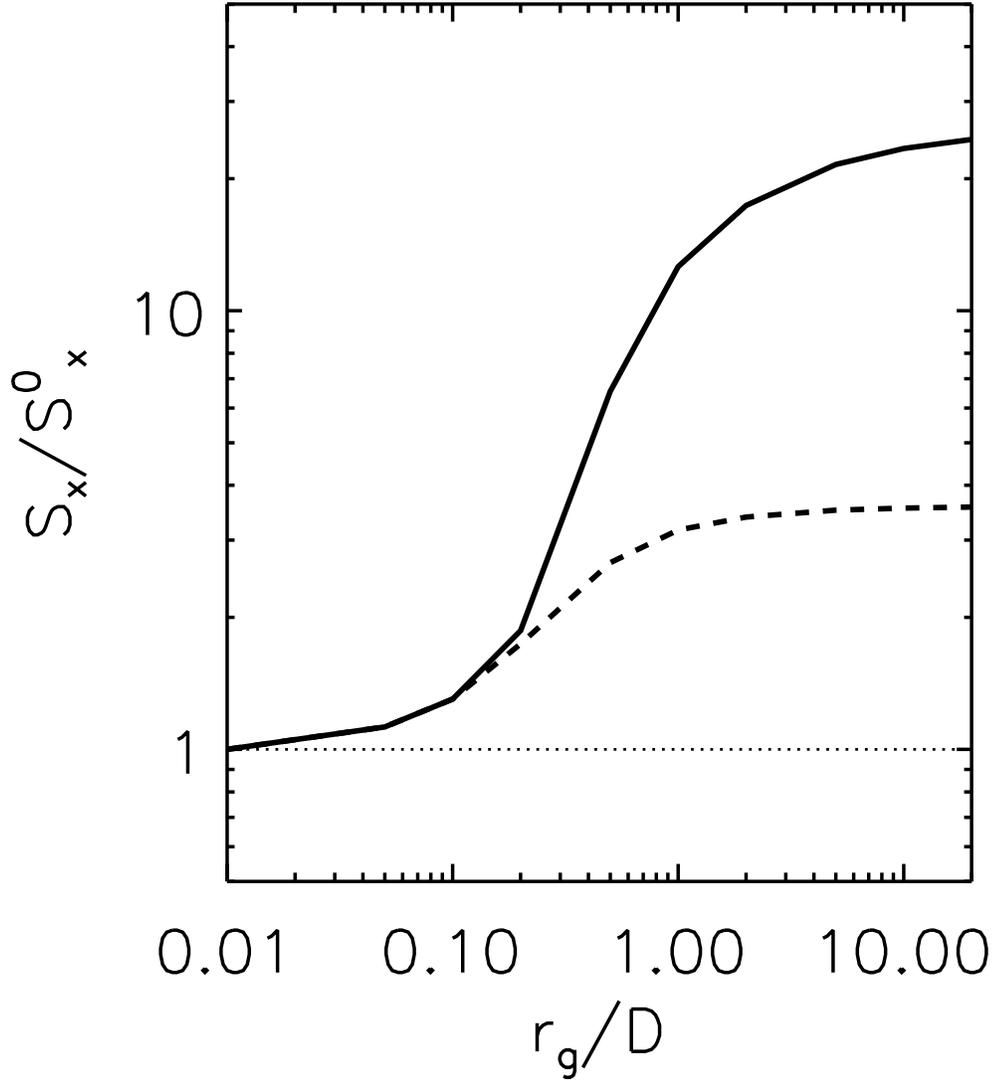}
\caption{The increase of the streaming $S_x$ in the $x$ direction relative to
the nominal streaming obtained for a uniform field.
The solid curve refers to equidistant alternating sectors, the dashed curve
refers to an asymmetric case with a dominant sector (see text).
\label{fig_kj1}}
\end{figure}

To represent a large scale gradient, $G_x$, we assume a solution of the
Boltzmann equation (1) in the form:
\begin{equation}
f(x,\varphi)=G_x x+F(x,\varphi),
\end{equation}
where $F(x,\varphi )$ is  periodic in $x$ and is required to be continuous 
at the currents sheets.
Integrating Eq. (1) immediately yields a physically plausible result that
streaming in the $x$ direction, $S_x$, must be constant.
The ratio of the computed  $S_x/G_x$ represents the effective diffusion
coefficient appropriate to this situation. 
We solve Eq. (1) numerically on a $200\times120$ grid in $x$ and $\varphi$.
Numerical results presented here were all obtained with $\lambda=5 r_g$
(i.e., $\Omega\tau=5$).
Furthermore, we consider a symmetric arrangement with $d_+=d_-=D/2$, as well as
an asymmetric one with a dominant sector by choosing $d_+=3D/4$ and $d_-=D/4$.

\begin{figure}
\plottwo{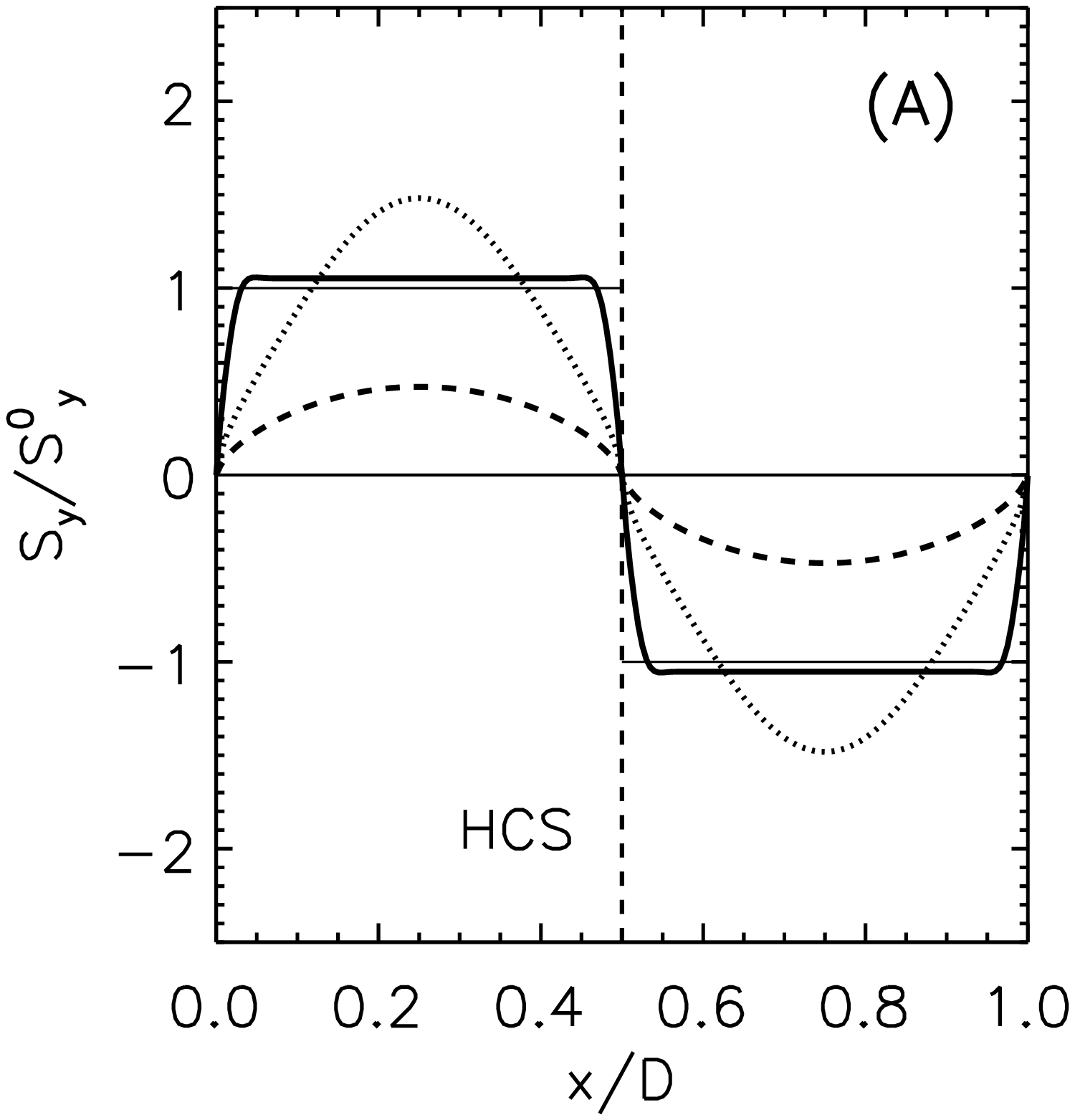}{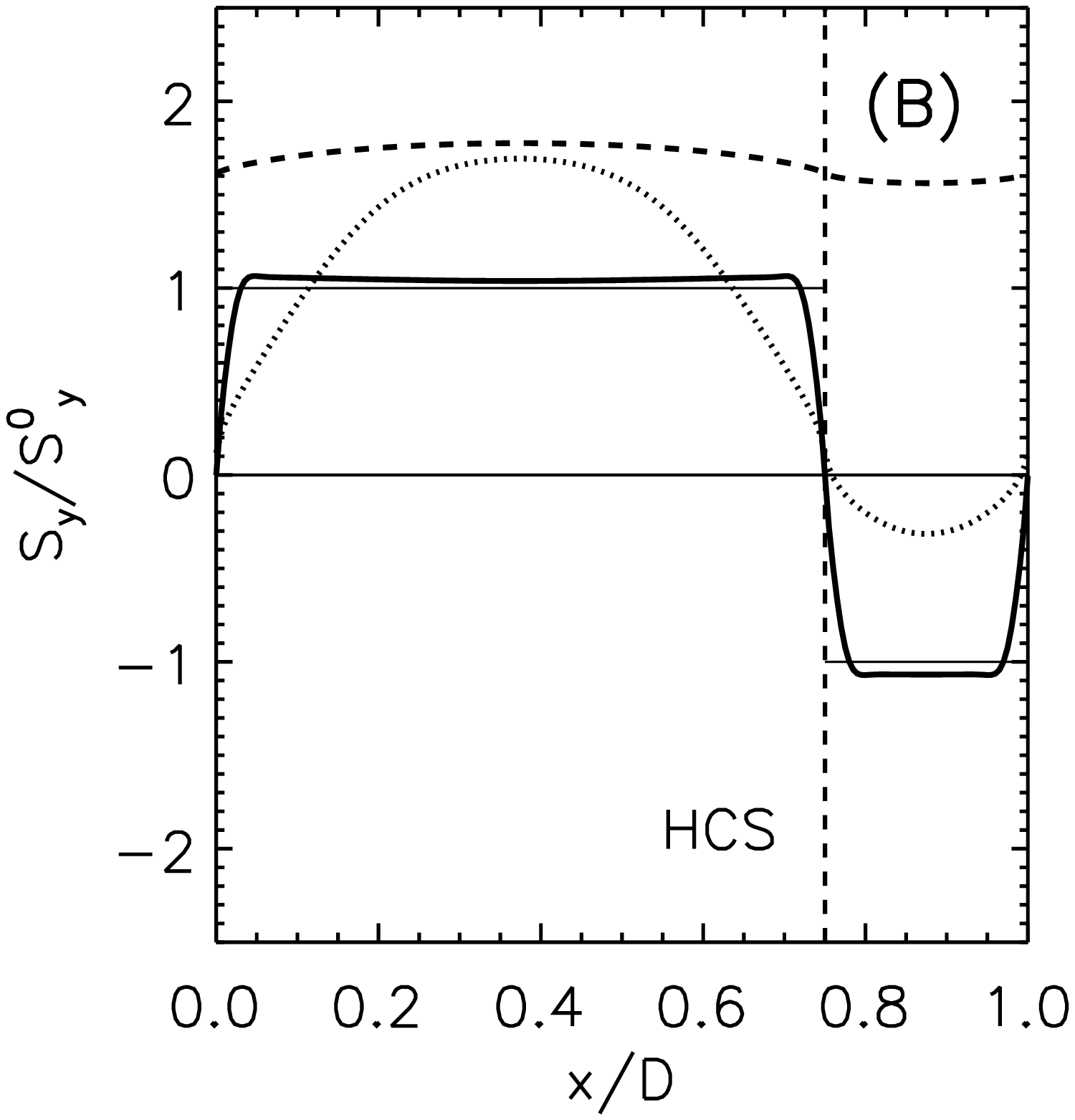}
\caption{Spatial variation of the diffusive streaming $S_y$ (solid:
$r_g/D=0.02$, dotted: $r_g/D=0.2$, dashed: $r_g/D=2$).
While $S_y(x)$ has a sharp transition at the HCS for $r_g/D \ll 1$, it decreases
in magnitude as $r_g/D$ increases. 
The left panel shows a symmetric case with two equally wide sectors, the right
panel shows the asymmetric arrangement with a dominant sector (see text).
Note that $S_y$ remains large in the latter case.
\label{fig_kj2}}
\end{figure}

Figure \ref{fig_kj1} shows how the streaming $S_x$ changes if the sectors become 
narrower, i.e., $r_g/D$ increases.
Because $S_x\sim\kappa_{xx}$ (assuming a constant spatial gradient), the figure
also shows the ratio between the cross-field diffusion coefficients with and
without the sectors.
The two curves refer to the symmetric and asymmetric arrangements, respectively.
Clearly, particle transport is significantly faster as the thickness of the
sectors becomes comparable or smaller than the particle's gyro-radius, $r_g$.
The increase is larger in the symmetric case, when $S_x$ approaches the 
limiting case obtained without any background field ($\Omega=0$), which is about
a 25-fold increase for our parameters ($\Omega\tau=5$).
For the asymmetric case the increase is slower.
Intuitively one can see that particles will sense the average $\Omega$, which
implies a twofold decrease in $\langle\Omega\rangle$, and a corresponding
fourfold increase in $S_x$ relative to the unipolar case.

Figure \ref{fig_kj2} illustrates the spatial variation of the $S_y$ streaming 
component for the values of $r_g/D=0.02$, $r_g/D=0.2$, and $r_g/D=2$ within a
single pair of sectors (the complete solution is of course periodic in $x$).
For small $r_g/D$ the solution is close to the formal discontinuous solution,
with a sharp transition occurring near the current sheet.
We note that the directional distribution has significant higher order
anisotropies in this region.
$S_y$ decreases as $r_g/D$ increases.
The right panel of Figure \ref{fig_kj2} shows the asymmetric case, with one
sector dominating.
This case yields a remarkably large streaming in the $y$-direction in both
sectors for $r_g/D=0.2$.
This seemingly counter-intuitive result can be readily understood recalling that
multiplying Eq. (1) by $\sin\varphi$ and averaging over $x$ gives 
\begin{equation}
{\langle S_y\rangle}=\langle\Omega\tau\rangle S_x,
\end{equation} 
which means that large streaming in the $x$ direction implies large streaming in
the $y$ direction as well, if one sector dominates.

\section{Test particle simulations of neutral sheet diffusion}
Building upon the conceptual random sector thickness model introduced in
\citet{florinski11}, we now proceed to investigate the transport of energetic
particles across a more realistic representation of magnetic sectors in the
heliosheath using a test-particle computer code for orbit integration.
Measured plasma velocities are small in the distant heliosheath
\citep{krimigis11}, and we ignore the background flow altogether.
Inside the sectors we assume that the magnetic field $\mathbf{B}$ is constant in
magnitude and points either in the $+\hat\mathbf{z}$ or the $-\hat\mathbf{z}$
directions, corresponding to the primarily azimuthal heliosheath magnetic field.
The motion of a charged particle is unrestricted along the field, and we only
consider their trajectories in the $xy$ plane (in other words, $z$ is an
ignorable coordinate in the present model).

In the computer model used here particles are injected into a two-dimensional
rectangular simulations box of size $L_x\times L_y$ containing a few hundred
alternating polarity sectors separated by current sheets (which represent
different segments of the same HCS).
Whereas a theoretical number of sectors in one 11-year solar cycle is about 150,
we use a larger number in the model to obtain a correct asymptotic behavior for
large times.
The current sheets are, on average, parallel to the $yz$ plane, and their mean
spacing along the $x$ (radial) direction is $d$.
We introduce perturbations to the form ($x$-displacement) of each current sheet
according to
\begin{equation}
x_s(y)=x_{s0}+\sum_{n=1}^{N_k} A_n\cos(k_n y+\beta_{sn}),
\end{equation}
where $x_{s0}=s d$ ($s$ is the integer label enumerating the sheets), $k_n$ is
the wavenumber of mode $n$, $\beta_{sn}$ is the random phase, and $A_n$ is the
amplitude, related to the power spectrum of the fluctuations $S(k)$ as
\citep[e.g.][]{giacalone99}
\begin{equation}
A_n^2=\frac{S(k_n)\Delta k_n\sigma_x^2}{\sum_{m=1}^{N_k} S(k_m)\Delta k_m}.
\end{equation}

The model one-dimensional spectrum of current sheet displacements consists of a
flat energy range between $k_\mathrm{min}$ and $k_b$, the ``bend-over'' wave
number, followed by an inertial range with $S\sim(k/k_b)^{-5/3}$ extending to
$k_\mathrm{max}$.
We use $N_k=150$ logarithmically equidistant modes separated by $\Delta k_n$.
For the heliosheath we assume that $k_\mathrm{max}=2\pi/0.05\,\mathrm{AU}^{-1}$
and $k_\mathrm{min}=2\pi/79\,\mathrm{AU}^{-1}$, a difference of 3.2 orders of
magnitude, with $k_b$ at the midpoint of the logarithmic interval.
This is our ``standard'' sector boundary displacement structure.
From the central limit theorem one may infer that the sector width distribution
thus obtained should be Gaussian, with mean $d$ and dispersion $\sigma_x$.
A direct numerical calculation of the distribution function confirmed that this
is indeed the case.

\begin{figure}
\plotone{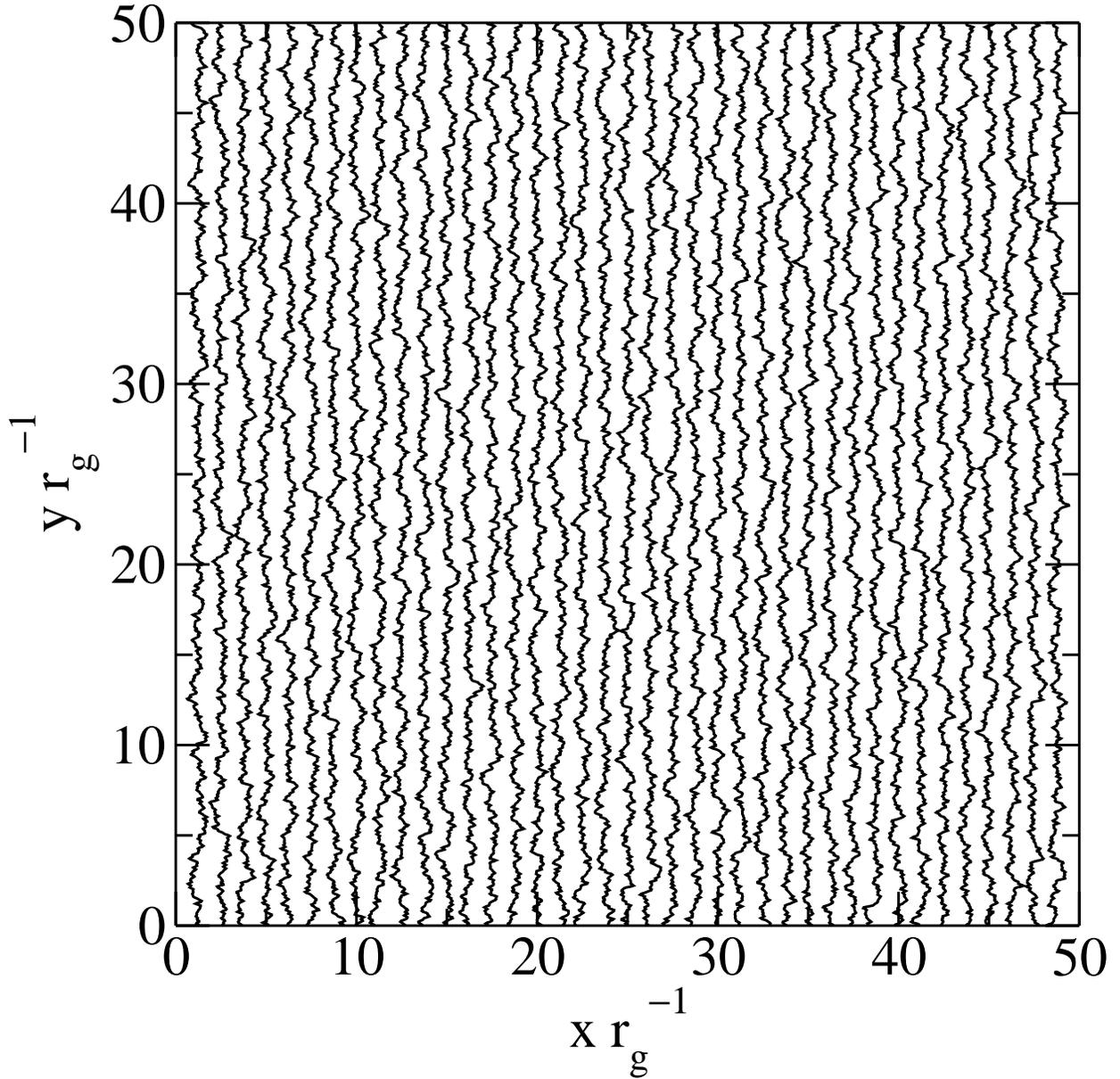}
\caption{Magnetic sector boundaries (a small fragment of the entire simulation)
in the $xy$ plane with $d=1.25r_g$, $\sigma_x=0.3r_g$, and
$k_b=1.57r_g^{-1}$.
The magnetic field is perpendicular to the page.
\label{fig_sectors}}
\end{figure}

In our simulations length is measured in units of the largest (for pitch angle
$\pi/2$) particle gyro-radius $r_g=pc/(ZeB)$, where $p$ is momentum and $Z$ is
its charge number.
The field is computed on a fine grid with equal mesh spacing in both directions;
a typical simulation may contain up to 30 billion grid cells.
The $L_x:L_y$ aspect ratio varies depending on the ratio of calculated diffusion
coefficients in the two directions, $\kappa_{xx}$ and $\kappa_{yy}$ (see below).
Typically we set $L_x/L_y\approx(\kappa_{xx}/\kappa_{yy})^{1/2}$.
The size of the box is chosen such that no particles escape through any side of
the box by the end of the run time.
Note that particles traveling the farthest from their starting point contribute
the most to diffusive transport statistics at later times.
If such particles were allowed to leave the system, the resulting diffusion
coefficients would be underestimated.
Figure \ref{fig_sectors} shows a fragment of the sector boundary structure
calculated with $d=1.25r_g$ and $\sigma_x=0.3r_g$.
The wiggles in the shape of the sheet are assumed to be a result of turbulent
flows in the solar wind and heliosheath that displace the HCS radially.

In our model the sector structure is controlled by three parameters: $d$,
$\sigma_x$, and $k_b$ (assuming the ratio $k_\mathrm{max}/k_\mathrm{min}$ is a
constant).
The resulting magnetic field topology may be viewed as a special case of strong
``turbulence'' with zero mean field.
This type of turbulence does not belong to the conventional slab/2D
classification system \citep{matthaeus90, bieber96}.
In our case  the fluctuating field $\delta\mathbf{B}\parallel\hat\mathbf{z}$ is
orthogonal to the plane spanned by the wavevectors (the $xy$ plane).
The turbulence is clearly non-axisymmetric; the perpendicular ($x$) power
spectrum is dominated by a signal at $\pi/d$ (twice the mean current sheet
spacing), and the parallel ($y$) spectrum has no clearly identifiable dominant
wavenumber.

Charged particle orbits are computed by integrating the Newton-Lorentz equation
of motion
\begin{equation}
\frac{d\mathbf{v}}{dt}=\mathbf{v}\times\mathbf{\Omega},
\end{equation}
where $\mathbf{v}$ is the particle's velocity and
$\mathbf{\Omega}=Ze\mathbf{B}/(\gamma mc)$ is its cyclotron frequency,
$m$ being the mass and $\gamma$ the relativistic factor.
Eq. (7) is solved numerically using the standard fourth-order Runge-Kutta
(RK4) method.
To achieve good accuracy we used a small time step, typically 
$\Delta t=2\pi(1000\Omega)^{-1}$.
The spatial resolution in both ($x$, $y$) directions was
$\Delta x=\Delta y=0.025r_g$ (dependent on particle's energy).

\begin{figure}
\plotone{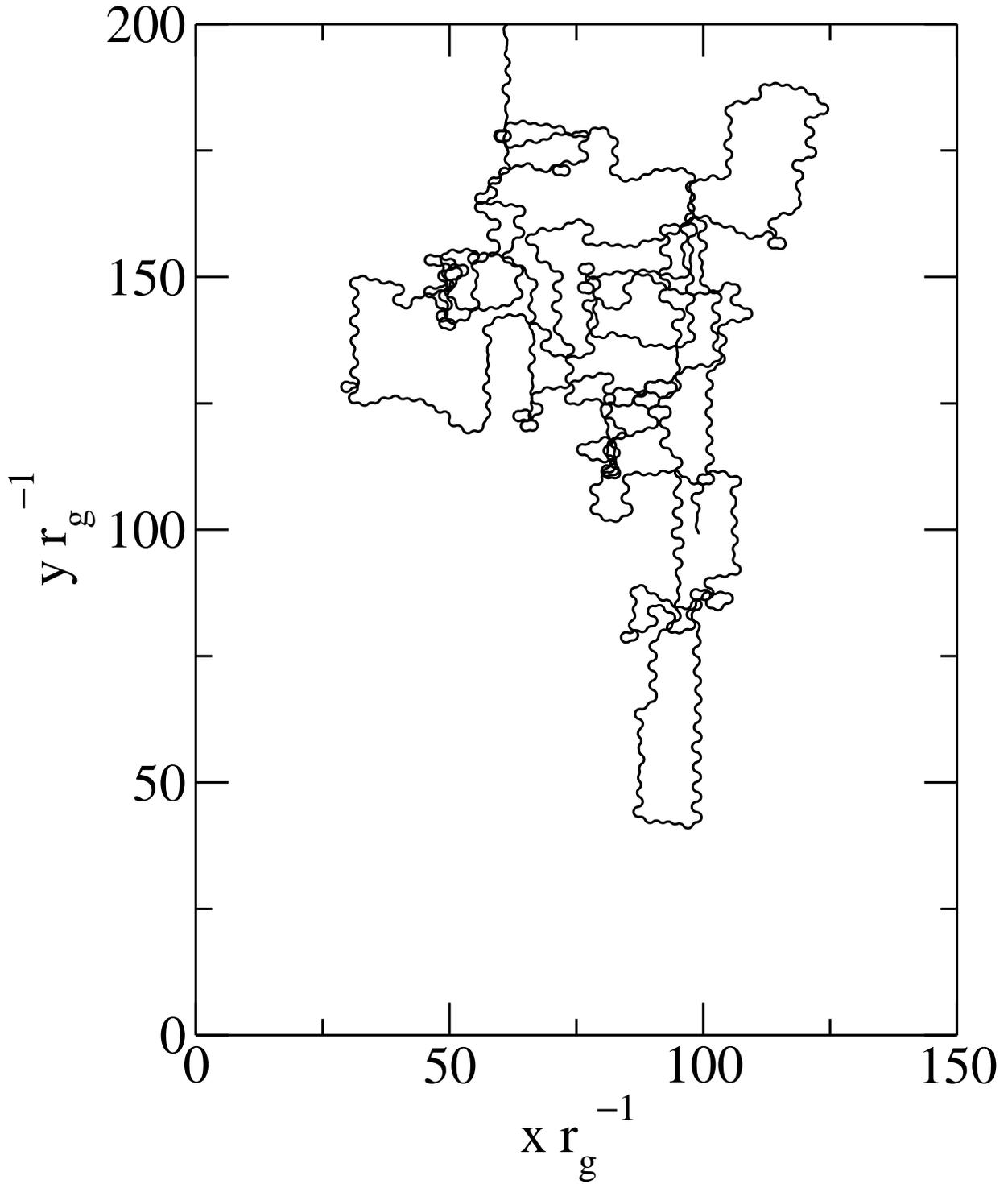}
\caption{A sample charged particle trajectory in a sectored magnetic field with
$d=1.25 r_g$, $\sigma_x=0.3r_g$, and $k_{\mathrm{min}}=1.57r_g^{-1}$,
illustrating parallel and perpendicular current sheet transport.
\label{fig_traj}}
\end{figure}

Figure \ref{fig_traj} illustrates typical particle behavior in a sectored
magnetic field with stochastically displaced boundaries.
The particle was injected with a fixed pitch angle $\theta=90^\circ$ near the
center of the box.
Notice the long excursions in the $y$ direction, corresponding to particle
drifting along a single current sheet for long periods of time.
The particle may subsequently transfer to an adjacent sheet when the two become
sufficiently close.
By examining Figure \ref{fig_traj} one can envision a certain parallel between
particle motion along the current sheets vs. motion across the sectors and
motion along and across the mean magnetic field in a conventional slab/2D
composite turbulence.
Indeed, in the latter case a particle stays on the same field line for a long
time, but can at times switch to a different field line as a result of
pitch-angle scattering events randomizing its gyrophase.
A constriction in the sector structure acts similar to a scattering center
allowing a particle to transfer from one current sheet to another.

\begin{figure}
\plotone{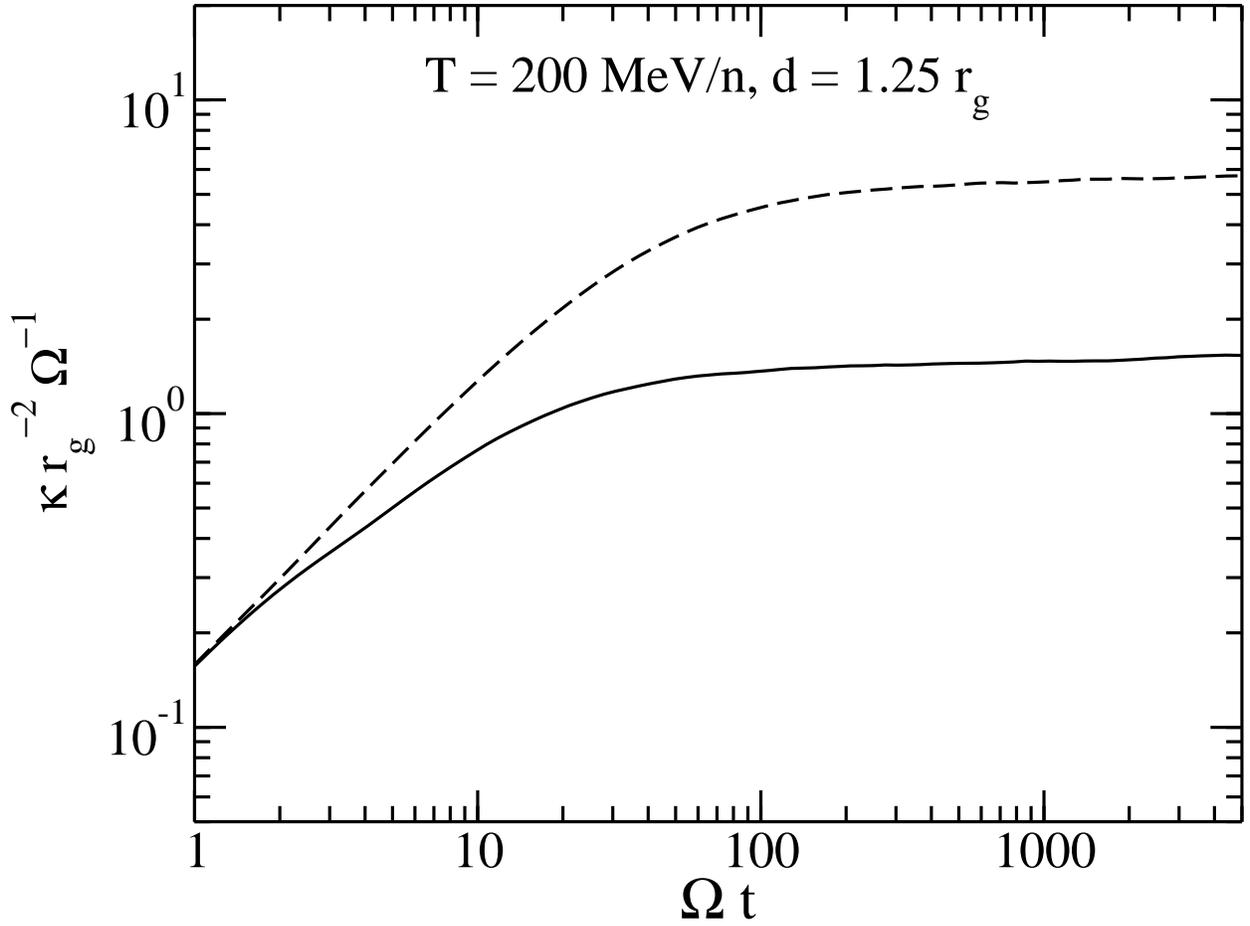}
\caption{Running diffusion coefficients, in units of $r_g^2\Omega$, as a
function of time.
Isotropically injected 200 MeV/n helium ions in a $4\,\mu$G magnetic field,
mean sector width $d=0.062$ AU.
\label{fig_rundif}}
\end{figure}

From our simulations it emerges that particle transport in both directions
becomes diffusive after a certain period of time (dependent on particle's
energy, mean sector width, and sector displacement properties).
We use running diffusion coefficients calculated as
\begin{equation}
\kappa_{xx}(t)=\frac{\langle[x(t)-x_0]^2\rangle}{2t},\quad
\kappa_{yy}(t)=\frac{\langle[y(t)-y_0]^2\rangle}{2t},
\end{equation}
where $(x_0, y_0)$ are the initial coordinates of the particles.
Here angular brackets mean ensemble averaging over both particles and field
realizations.
We typically use 25 sector structure realizations and 2000 particles per
realization injected in an isotropic shell distribution with a specified kinetic
energy $T$.
Figure \ref{fig_rundif} shows $\kappa_{xx}$ and $\kappa_{yy}$ as functions of time
for a demonstration run for isotropically  injected 200 MeV/n He$^{2+}$ in a
field with a strength $B=0.4$ nT, and the sheet structure as in Fig.
\ref{fig_sectors}.
The size of the simulation box was 107 AU by 573 AU in the $x$ and $y$
directions, respectively.
Such a large size was chosen to ensure that we measure correct asymptotic
values of the running diffusion coefficients (see discussion above).
All 50,000 particles remained inside the box after 2000 gyro-periods.

After a brief superdiffusive phase, lasting a few tens of gyro-periods, when
particles are streaming primarily along individual neutral sheets (marked by
increasing $\kappa(t)$), normal diffusion is established in both directions.
Asymptotic values of $\kappa_{xx}$ and $\kappa_{yy}$ are obtained from this late
phase of the particles' trajectories.
For the present choice of parameters, the cross-sector diffusion coefficient
is a factor of $\sim 4$ smaller than the ``parallel'' diffusion coefficient.
The ratio $\kappa_{xx}/\kappa_{yy}$ depends sensitively on the width of the
sectors.
The next section demonstrates the dependence of diffusion coefficients on
particle's energy and on the properties of current-sheet ``turbulence''.

\section{Dependence on sector properties and particle energy}

We first discuss the dependence of diffusion coefficients on energy.
In the following numerical experiments the current sheets had the same
properties using the ``standard'' choice of parameters $B$, $d$, $\sigma_x$,
$k_\mathrm{min}$, $k_b$, and $k_\mathrm{max}$ as described in the previous
section.
The energy of He$^{2+}$ ions was varied between $10$ MeV/n and $1$ GeV/n.
The size of the simulation box was adjusted accordingly to prevent particles
from leaving.

\begin{figure}
\plotone{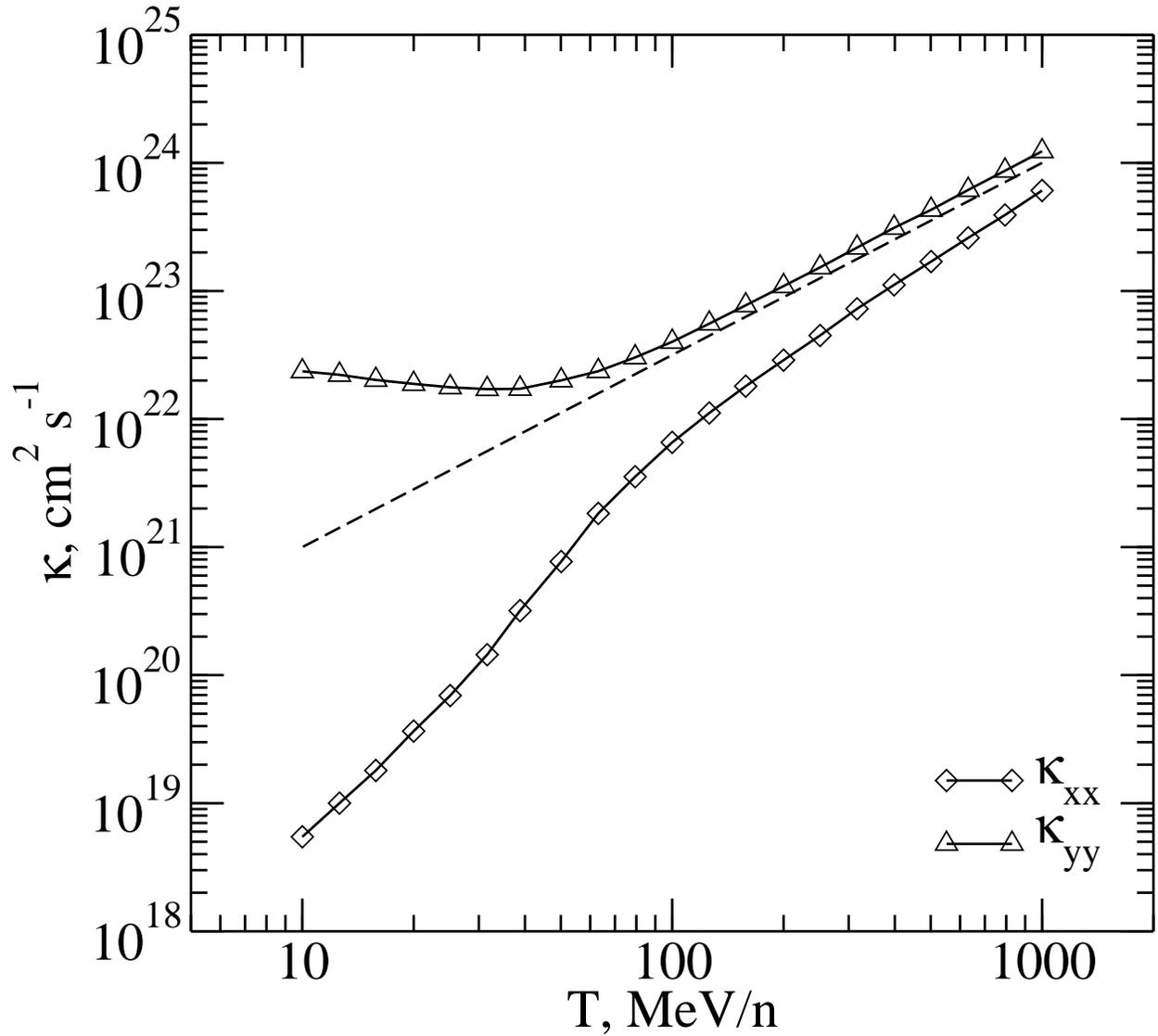}
\caption{Parallel ($\kappa_{yy}$) and perpendicular ($\kappa_{xx}$) diffusion
coefficients for galactic He as a function of energy per nucleon.
The dashed line has a power law slope of $3/2$.
\label{fig_kpkpe}}
\end{figure}

The results of this experiment are shown in Figure \ref{fig_kpkpe}.
At the lowest energy (10 MeV/n) the Larmor radius is about 4 times smaller than
the mean distance between the sheets and perpendicular diffusion is strongly
suppressed.
Parallel transport still appears to be super-diffusive by the end of the
simulation, which may explain the upturn in $\kappa_{yy}$ at low energies.
By about 100 MeV/n the neutral sheet separation is about equal the Larmor radius
leading to a dramatic increase in $\kappa_{xx}$.
Here transport is strictly diffusive and both coefficients increase with ion
energy.
The parallel diffusion coefficient is well approximated by a power-law
with an index of $3/2$, while the perpendicular coefficient increases somewhat
faster as $\sim T^2$.
In the non-relativistic regime, this behavior is comparable with a quasi-linear
theory result $\kappa_{yy}\sim T^{1+\alpha/2}$ for a turbulent power-law
spectral slope $\alpha=1$ \citep[e.g.,][]{jokipii71}.
It is not clear whether there is a connection between the two or it is a simple
coincidence.
Note that in our model the spectrum of fluctuations in the current sheets
varies between $-5/3$ and zero.

\begin{figure}
\plotone{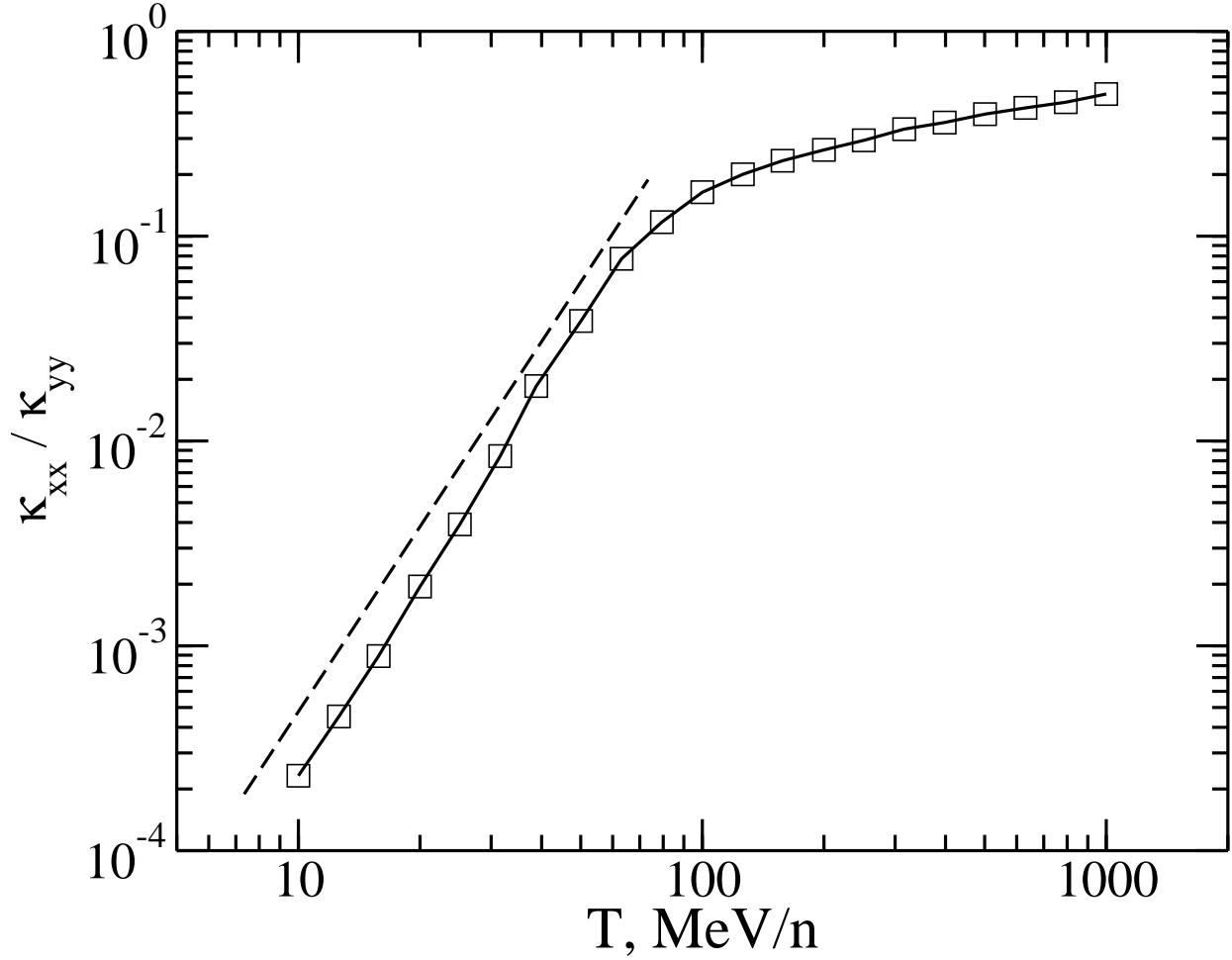}
\caption{Ratio of perpendicular ($\kappa_{xx}$) to parallel ($\kappa_{yy}$)
diffusion coefficients for galactic He as a function of energy per nucleon.
The dashed line has a slope of 3.
\label{fig_kratio}}
\end{figure}

Figure \ref{fig_kratio} shows the dependence of $\kappa_{xx}/\kappa_{yy}$ on
ion energy.
The ratio is a rapidly increasing function of energy ($\sim T^3$ at low energy)
and approaches a unity (isotropic diffusion) at high energy.
The transition from slow to fast perpendicular diffusion occurs at
$r_g\simeq d$.

In the second set of numerical experiments we simulated particles at two fixed
energies, 50 MeV/n and 250 MeV/n, while varying the distance $d$ between the
sheets and the displacement magnitude $\sigma_x$ while keeping the ratio
$\sigma_x/d$ a constant ($=0.24$).
This approximates a situation where the stack of sectors is compressed, but the
wiggles become smaller as well (the complete structure is flattened on approach
to the stagnation point).
The magnitude of $\mathbf{B}$ was kept a constant implying that the excess field
was removed from the region by a latitudinal flow.
The current sheet turbulence spectrum was again of the ``standard'' type
(described following Eq. (6)).

\begin{figure}
\plotone{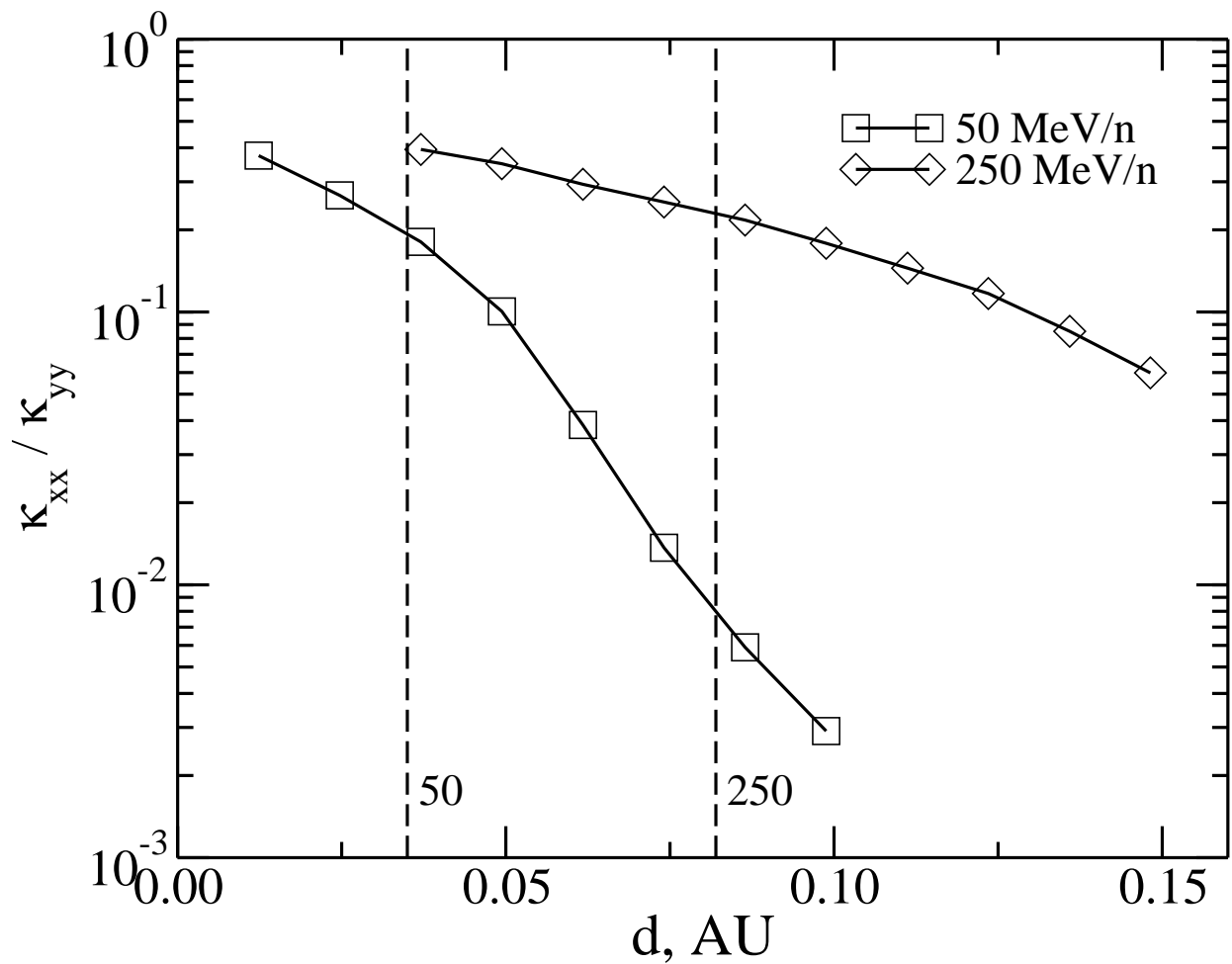}
\caption{Ratio of perpendicular ($\kappa_{xx}$) to parallel ($\kappa_{yy}$)
diffusion coefficients at 50 MeV/n (squares) and 250 MeV/n (diamonds) as a
function of mean sector thickness.
The ratio of the magnitude of sector boundary fluctuations and the sector
thickness was kept a constant (0.24).
Vertical dashed lines are the gyro-radii of both ions in a 0.4 nT magnetic
field.
\label{fig_kpkpds}}
\end{figure}

We show the ratio $\kappa_{xx}/\kappa_{yy}$ in Figure \ref{fig_kpkpds}.
For 50 MeV/n ions the ratio drops rapidly as the distance $d$ becomes larger
than the gyro-radius.
Conversely, at 250 MeV/n the decrease is much more gradual.
The main reason for this difference is the fact that the power spectrum of
neutral sheet oscillations is the same in both simulations (with 50 MeV/n and
250 MeV/n).
The two would be alike only if the neutral sheet displacement wavelengths were
also varied $\sim d$.

In the final experiment we vary the turbulence spectrum for a fixed energy
(100 MeV/n) and sector width ($d=1.25 r_g$).
Instead of a broadband spectrum employed in the previous simulations, here we
use a narrow-band spectrum with a constant power spectral density between
$k_\mathrm{min}$ and $k_\mathrm{max}$.
The spectrum extends a single decade in wavenumber and is represented with
$N_k=50$ modes.
Running diffusion coefficients at the end of each run are recorded as the
asymptotic values.

\begin{figure}
\plotone{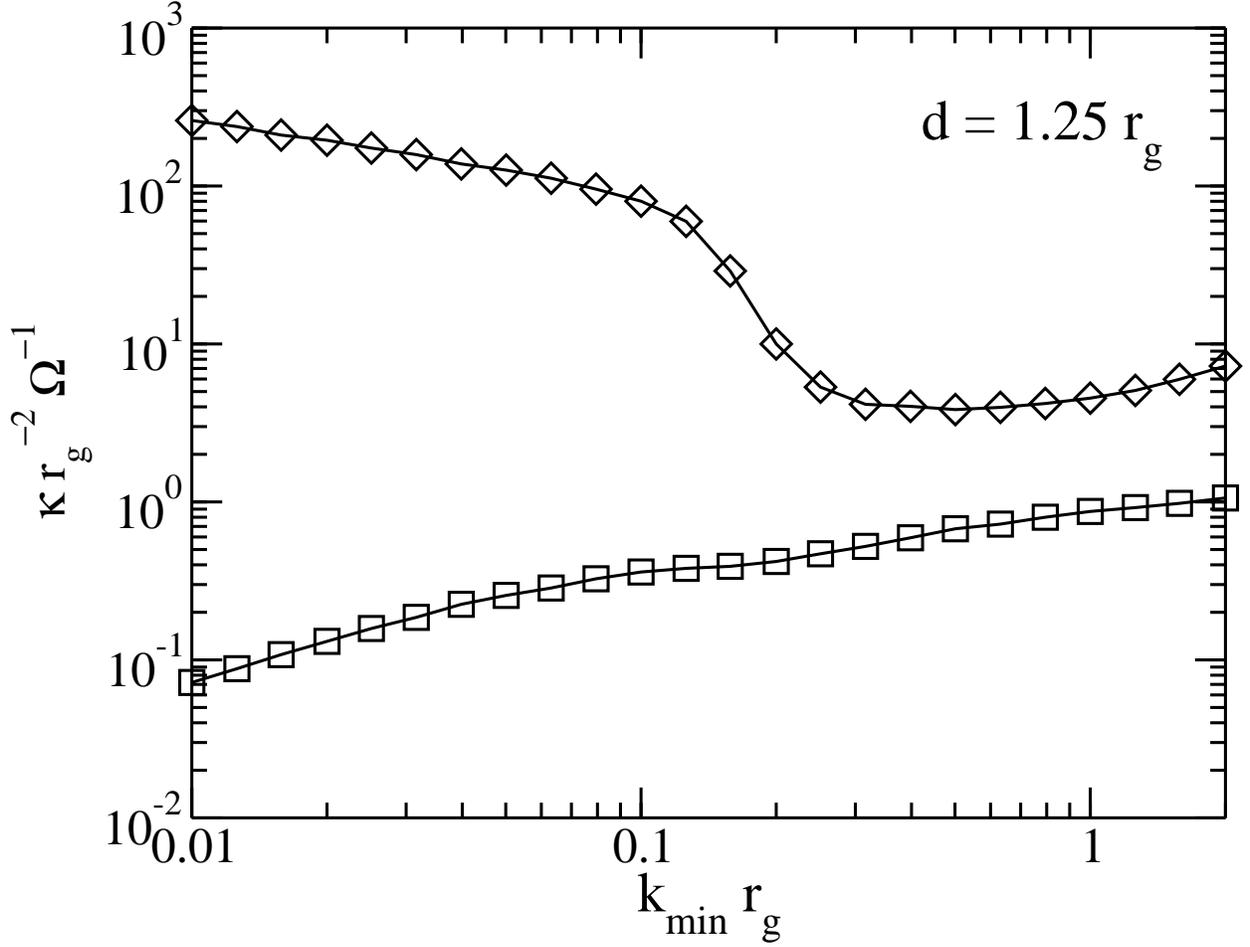}
\caption{Parallel ($\kappa_{yy}$, diamonds) and perpendicular ($\kappa_{xx}$,
squares) diffusion coefficients of 100 MeV/n He ions as a function of 
$k_\mathrm{min}$.
The wavenumber and the diffusion coefficients are normalized to $r_g$ and
$r_g^2\Omega$, respectively.
The mean sector width $d=1.25 r_g$.
\label{fig_wave1}}
\end{figure}

The results are plotted in Figure \ref{fig_wave1}.
We see that increasing the wavelength leads to an increase in $\kappa_{yy}$ and
a decrease in $\kappa_{xx}$.
The latter is easily understood from the fact that by reducing $k_\mathrm{min}$
the mean distance between neutral sheet constrictions, acting as
scattering centers, increases leading to a reduction in scattering rates.
There is a rapid drop in $\kappa_{yy}$ near what could be described as
``resonant'' wavenumber where $k r_g$ is of order 1.
It is not clear what causes the upturn at large $k_\mathrm{min}$.
It is conceivable that the model is unable to resolve such fine structures with
good accuracy.

\section{Discussion}
Analytic and numerical results presented here clearly demonstrate that a warped
HCS can facilitate charged particle transport perpendicular to the mean magnetic
field, if the distance between the folds $d$ is comparable to (or smaller than)
the particle's Larmor radius, the condition that is expected to be satisfied
within about 10 AU before the heliopause.
The analytic treatment of the problem was performed using the Boltzmann equation
for a non-gyrotropic velocity distribution.
Unlike the numerical model, the sector width was constant.
Instead, stochasticity was provided by small-angle gyrophase scattering events.
Our results show a steady increase in the cosmic-ray differential current
perpendicular to the sectors (and therefore, in the cross-field diffusion
coefficient) with the ratio $r_g/d$.
Gyration can produce net streaming along the sector planes if the widths of the
positive and the negative sectors are different.
The magnitude of the streaming flux tends to increase with $r_g/d$.

In our test-particle simulations we found that particles undergo diffusive
random walk in the plane perpendicular to $\mathbf{B}$, i.e., in the radial
and latitudinal directions.
The efficiency of this process depends critically on the size of neutral sheet
``wiggles'' and the mean thickness of the alternating polarity sectors.
The ratio of perpendicular and parallel diffusion coefficients is zero in the
limit of very long-wavelenfth fluctuations, and when the ratio $r_g/d$ is small.
In the opposite limit (short-wavelength fluctuations, $r_g/d>1$) the ratio
approaches unity.
In the latter regime, the value of $\kappa_{rr}$ could be as large as
$6\times10^{21}-6\times10^{23}$ cm$^2$/s for He$^{2+}$ ions with energies
between 100 MeV and 1 GeV per nucleon, assuming a magnetic field strength of
0.4 nT (see Figure \ref{fig_kpkpe}).
These values are larger than the perpendicular diffusion coefficients calculated
using the standard theory for solar-wind fluctuations.
For example, in the model of \citet{florinski09} $\kappa_{\perp}$ is between
$4\times10^{21}$ and $4\times10^{22}$ cm$^2$/s for 100 MeV and 1 GeV protons,
respectively.
For helium with the same energy per nucleon the diffusion coefficients will be
larger by up to a factor of 4 (for resonant interactions in the energy range of
the turbulence with $\kappa_{\parallel}\sim r_g^2$ and
$\kappa_{\perp}\sim\kappa_{\parallel}$).
The current-sheet radial diffusion coefficient is about a factor of 5 larger.
Remember, however, that conventional scattering by turbulence was not included
in our model.
We expect our results to be more reliable in a situation where the level of
background turbulent activity at resonant wavenumbers is relatively low.

The interplay between the sheet-parallel and perpendicular transport in our
model resembles in several ways conventional diffusion along and across the
magnetic field.
Constrictions between the adjacent sheets appear to play the role of scattering
centers.
Nevertheless, one should exercise caution before drawing parallels between the
two.
Parallel diffusion in QLT is a resonant process, which is clearly not the case
for current sheet diffusion.

The current sheet transport mechanism is most efficient for galactic ions with
energies of the order 100 MeV per nucleon and above.
Voyager 1 observations of galactic helium made near the end of 2011 at a
heliocentric distance of $\sim 119$ AU (McDonald et al., AGU abstract) are
close or in excess of the predicted interstellar intensities from
\citet{webber09}.
If Voyager 1 is 10 AU in front of the heliopause, the current-sheet effect could
be responsible for the lack of modulation.
Another possible cause is the absence of a radial plasma flow \citep{krimigis11}
which would otherwise expel energetic particles from the heliosheath.

The diffusion mechanism discussed here may facilitate the transport of
interstellar dust grains with size $<0.01\;\mu$m that have a gyroradius similar
to that of a few hundred MeV cosmic-ray ion \citep{czechowski03}.
Conversely, it is not efficient for low-rigidity particles such as electrons,
because their gyro-radius is much smaller than the distance between the adjacent
sheets.
The recent rise of the low-energy electrons measured by the Voyagers
\citep{caballerolopez10, webber12}, is probably not directly related to the
current sheet diffusion effect discussed here.
At such small rigidities the effect of magnetic field meandering is likely to
take precedence over current sheet transport as described by our current model.
The model is somewhat restricted by the assumption that the direction along
$\mathbf{B}$ is an ignorable coordinate; this could lead to an under-estimate
of the diffusion rate in the radial direction.

We realize that a more accurate diffusion model should include ``conventional''
turbulence (variations in the magnitude and direction of $\mathbf{B}$) in
addition to the current-sheet oscillations.
Such a project would require massively larger computer simulations taking
thousands of CPU-hours for a single run, but may be ultimately needed solve the
problem of cosmic-ray transport in the distant heliosheath.
Despite technical difficulties, in the future it might be possible to combine
the two kinds of diffusion into a single three-dimensional model of energetic
charged particle transport.

\acknowledgments
V.F. was supported, in part, by NASA grant NNX10AE46G, NSF grant AGS-0955700
(also supporting X.G.) and by a cooperative agreement with NASA Marshall Space
Flight Center.
J.K. acknowledges support by NASA grant NNX08AQ14G.


\end{document}